\documentstyle{article}
\newcommand{\lascia}[1]{}
\newcommand{\mb}[1]{\mbox{\normalsize\boldmath $#1$}}
\newcommand{\md}[1]{\langle #1\rangle}
\newcommand{\Dsl}{D\hspace{-6.6pt}{/}\,}
\def\circa#1{\,\raise.3ex\hbox{$#1$\kern-.75em\lower1ex\hbox{$\sim$}}\,}
\def\SU{{\rm SU}}

\newcommand{\tfrac}[2]{{\textstyle{\frac{#1}{#2}}}}

\newcommand{\hM}{\hat{M}}
\newcount\Mac  \Mac=0  
\newcommand{\ifMac}[2]{\ifnum\Mac=1 #1 \else #2 \fi}
\def\Tr{\mathop{\rm Tr}}
\def\One{\hbox{1\kern-.24em I}}
\newcommand{\MSbar}{{\overline{\hbox{\sc ms}}}}
\newcommand{\DRbar}{{\overline{\hbox{\sc dr}}}}
\newcommand{\Li}{\hbox{Li}_2}
\newcommand{\mub}{\bar{\mu}}

\newcommand{\GeV}{\,{\rm GeV}}
\newcommand{\TeV}{\,{\rm TeV}}
\newcommand\Ord{{\cal O}}

\newcommand{\epsIR}{\varepsilon_{\rm ir}}
\newcommand{\epsUV}{\varepsilon_{\rm uv}}
\newcommand{\eps}{\varepsilon}
\newcommand{\hc}{\hbox{h.c.}}
\newcommand{\eq}[1]{~{\rm (\ref{eq:#1})}}

\def\Red{}
\def\Black{}
\def\Blue{}

\newcommand{\NP}{Nucl. Phys.}
\newcommand{\PRL}{Phys. Rev. Lett.}
\newcommand{\PL}{Phys. Lett.}
\newcommand{\PR}{Phys. Rev.}
\makeatletter
\def\art{\@ifnextchar[{\eart}{\oart}}
\def\eart[#1]#2#3#4#5#6{{\rm #2}, {\em #3 \bf #4} {\rm (#6) #5}}
\def\hepart[#1]#2{{\rm #2, \em#1}}
\newcommand{\oart}[5]{{\rm #1}, {\em #2 \bf #3} {\rm (#5) #4}}
\newcommand{\y}{{\rm and} }

\newcounter{alphaequation}[equation]
\def\thealphaequation{\theequation\hbox to
0.6em{\hfil\alph{alphaequation}\hfil}}
\def\eqnsystem#1{
\def\@eqnnum{{\rm (\thealphaequation)}}
\def\@@eqncr{\let\@tempa\relax \ifcase\@eqcnt \def\@tempa{& & &} \or
  \def\@tempa{& &}\or \def\@tempa{&}\fi\@tempa
  \if@eqnsw\@eqnnum\refstepcounter{alphaequation}\fi
\global\@eqnswtrue\global\@eqcnt=0\cr}
\refstepcounter{equation} \let\@currentlabel\theequation \def\@tempb{#1}
\ifx\@tempb\empty\else\label{#1}\fi
\refstepcounter{alphaequation}
\let\@currentlabel\thealphaequation
\global\@eqnswtrue\global\@eqcnt=0 \tabskip\@centering\let\\=\@eqncr
$$\halign to \displaywidth\bgroup \@eqnsel\hskip\@centering
$\displaystyle\tabskip\z@{##}$&\global\@eqcnt\@ne
\hskip2\arraycolsep\hfil${##}$\hfil& \global\@eqcnt\tw@\hskip2\arraycolsep
$\displaystyle\tabskip\z@{##}$\hfil
\tabskip\@centering&\llap{##}\tabskip\z@\cr}

\def\endeqnsystem{\@@eqncr\egroup$$\global\@ignoretrue} \makeatother

\begin{document}
\begin{quote}
{\em Feb 1998}\hfill {\bf IFUP--TH 63/97}\\
\phantom{.} \hfill{\bf hep-ph/9802446}
\end{quote}
\bigskip
\centerline{\huge\bf\Red Next-to-leading order corrections}
\centerline{\huge\bf\Red to gauge-mediated gaugino masses }
\bigskip\bigskip\Black
\centerline{\large\bf Marco Picariello {\rm and} Alessandro Strumia} \vspace{0.3cm}
\centerline{\em Dipartimento di Fisica, Universit\`a di Pisa and}
\centerline{\em INFN, sezione di Pisa,  I-56126 Pisa, Italia}\vspace{0.3cm}
\bigskip\bigskip\Blue
\centerline{\large\bf Abstract}
\begin{quote}\large\indent
We compute the next-to-leading order corrections to
the gaugino masses $M_i$ in gauge-mediated models
for generic values of the  messenger masses $M$
and discuss the predictions of unified messenger models.
If $M<100$~TeV there can be up to $10\%$ corrections to the
leading order relations $M_i\propto \alpha_i$.
If the messengers are heavier there are only few \%{} corrections.
We also study the messenger corrections to gauge coupling unification:
as a result of cancellations dictated by supersymmetry,
the predicted value of the strong coupling constant is typically only
negligibly increased.
\end{quote}\Black

\section{Introduction}
The ``{\em gauge mediation}'' scenario for supersymmetric
particle masses~\cite{GM} can be realized
in reasonable models~\cite{GMmodels,DDGR}.
Furthermore, with a unified spectrum
of messenger fields, it gives rise to some stable and acceptable prediction
for the spectrum of supersymmetric particles.
One of these predictions is the
`unification prediction' for the gaugino masses $M_i$ ($i=1,2,3$): at
one-loop order the RGE-invariant
ratio $\rho_i\equiv M_i/\alpha_i $ is the same for all the three factors
of the SM gauge group
$G_{\rm SM}=\bigotimes_i G_i={\rm U}(1)_Y\otimes \SU(2)_L\otimes \SU(3)_C$.
This prediction is sufficiently stable that it is interesting
to compute it with more accuracy.

A detailed computation allows a comparison with the gaugino spectrum predicted
by the alternative scenario known as ``{\em unified supergravity}''~\cite{SuGraSoft}.
Unification relations for the $\rho_i(E)$ are infact the more stable prediction
of this second scenario:
gauge couplings and gaugino masses could receive sizeable GUT threshold corrections;
but these corrections largely cancel out in the ratios
$\rho_i(M_{\rm GUT})$,
since $\alpha_i$ and $M_i$ have the same one-loop RGE evolution.
The testable predictions for the low-energy running $\rho_i$ ratios
in the $\DRbar$ scheme~\cite{DR} are
(including NLO RGE corrections~\cite{RGEM}, but neglecting possible unknown
$\Ord(\%)$  GUT-scale threshold effects~\cite{SoglieMi})
\begin{equation}
\frac{\rho_1(M_Z)}{\rho_2(M_Z)}\approx 1.02,\qquad
\frac{\rho_2(M_Z)}{\rho_3(M_Z)}\approx 0.97.
\end{equation}
This should be compared with
the corresponding prediction in gauge-mediation models for
the ratios $\rho_i/\rho_j$, plotted in fig.s~\ref{fig:rho10} and~\ref{fig:rhoEtc}.

\smallskip

The computation of gaugino masses with NLO precision is done
in sections~\ref{calcolo} and \ref{fenomenologia}
for generic values of the messenger mass $M$.
It requires the following main steps:
\begin{itemize}
\item[1.] compute the renormalized running gaugino masses, $M_i(E_H)$
at $E_H\sim M$,
in the effective theory without messengers.

\item[2.] compute the running gaugino masses, $M_i(E_L)$ at $E_L\sim M_Z$,
evolving $M_i(E_H)$ down to the Fermi scale with 2~loop RGE equations.

\item[3.] use $M_i(E_L)$ to compute directely measurable quantities,
like the gaugino
pole masses, $M_i^{\rm pole}$, including all 1~loop effects at the electroweak scale.

\end{itemize}
The computation necessary for step~1 is done in section~\ref{calcolo}.
We will employ supersymmetric dimensional regularization, so that
the renormalization scale $E$ will be the $\DRbar$ scale, $\mub$.
The RGE necessary for step~2 (recalled in appendix~B)
can be read from the literature~\cite{RGEM}.
The one-loop expressions for pole gaugino masses in terms of running parameters
are also well known~\cite{Mpole}.
Since various unmeasured and unpredicted parameters
(like the so-called $\mu$-term)
would enter the final step 3, we prefer to show our final predictions for the
running MSSM gaugino masses renormalized at $\mub_L=M_Z$,
without including the gauge corrections at the electroweak scale.

We will compute these predictions in {\em unified messenger models}.
One more step is necessary to impose
the unification constraints on the messenger spectrum, namely
\begin{itemize}
\item[0.] compute the messenger spectrum, $M_n(E_H)$,
evolving the unified $M_n(M_{\rm GUT})$ down to the messenger
scale $E_H\sim M_n$ with 2~loop RGE equations.
\end{itemize}
The necessary RGE equations are given in appendix~B.
In sec.~\ref{fenomenologia} we study the predictions
of gauge mediation models with an unified messenger spectrum.
If the messenger spectrum is only negligibly splitted
by supersymmetry breaking effects, the NLO corrections to
the LO unification-like relations $M_i\propto \alpha_i$ are
around few $\%$ and numerically not much different from the
ones present in unified supergravity models.
Larger effects (up to $10\%$) can be present if the messengers are
very light, $M\circa{<}50\TeV$.

We also study the corrections to gauge coupling unification
due to the presence of messenger fields below the unification scale.
Messenger threshold effects largely cancel messenger corrections to
two~loop RGE running, as dictated by supersymmetry~\cite{hol,RGEMhol}.

\section{Computation}\label{calcolo}
In this section we do the computations necessary for step~1.
We assume that the messenger fields sit in real representations
$R=\bigoplus_n R_n$ of the SM gauge group
and have a supersymmetric mass term $M_n$ together with a non-supersymmetric
mass term $F_n$.
For the moment we assume that the messengers lie
in self-conjugate complex representations, $R_n=X_n\oplus\bar{X}_n$.
See appendix~A for a more detailed discussions of the notations, of the model,
of its spectrum, and of the relevant Lagrangian.

In order to compute the running gaugino masses $M_i(\mub)$ at $\mub\circa{<}M_n$
in the effective theory we first
compute the gauge-independent {\em pole} gaugino masses in the two versions of the theory.
The computation in the full theory (with messengers)
is done in section~\ref{full} ---
the computation in the effective theory (with messengers integrated out)
is done in section~\ref{eff}.
Requiring that the two theories describe the same gaugino masses
up to second order in $\alpha_i$ we
get, in section~\ref{match} the gaugino mass terms in the effective theory.

\medskip

We employ the Feynman-Wess-Zumino gauge
and the supersymmetric $\DRbar$ regularization~\cite{DR}
in both versions of the theory.
The final $\DRbar$ values of the gauge couplings $\alpha_i$ and of the gaugino masses $M_i$
can be converted into the $\MSbar$ ones
(i.e.\ the ones obtained with na\"{\i}ve dimensional regularization)
using
$$\alpha_{\MSbar}=\alpha_{\DRbar}(1-\frac{C(G)}{3}\frac{\alpha}{4\pi}),\qquad
M_{\MSbar}=M_{\DRbar}(1+C(G)\frac{\alpha}{4\pi})$$
where the group factors are defined as follows.
For each representation $R$ of a gauge group $G=\bigotimes_i G_i$ we define
the ``Dinkin index'' $T_i(R)$ and the ``quadratic Casimir'' $C_i(R)$
in terms of the generators $T^a_{Ri}$ as
\begin{equation}
\sum_a T_{Ri}^a \cdot T_{Ri}^a = C_i(R) \One,\qquad
\Tr T^a_{Ri} T^b_{Rj}=T_i(R)\delta_{ij}\delta^{ab}.
\end{equation}
With generators canonically normalized so that $T(\mb{n})=T(\bar{\mb{n}})=1/2$
for the fundamental $\mb{n}$ representation of a $\SU(n)$ group,
the quadratic Casimir of the adjoint representation of a $\SU(n)$ group is
$C(G)=T(G)=n$, while $C(G)=0$ for $G$ a U(1) factor.
The values of the coefficients for the SM gauge group
and for representations 
contained in $5\oplus\bar{5}$ and $10\oplus\overline{10}$ representation of SU(5)
are given in table~\ref{tab:CT}.

\begin{table}
$$\begin{array}{|cl||cl|}
\multicolumn{2}{c}{\hbox{graph $\Gamma$ and its value $V^\Gamma$}}&
\multicolumn{2}{c}{\hbox{graph $\Gamma$ and its value $V^\Gamma$}} \\ \hline
XX& \alpha_j C_j(X) (-\frac{10}{3}) &
X\tilde{X}&\alpha_j [C_{j}(X)-\frac{1}{2}\delta_{ij}C_i(G)](2/\eps+4)\\
\widetilde{XX}&\alpha_j C_j(X) \frac{4}{3}(-1/\eps+1)&
\widetilde{X\tilde{X}}& \alpha_j [C_{j}(X)-\frac{1}{2}\delta_{ij}C_i(G)](-4)\\
\tilde{X}\tilde{X}&\alpha_j C_j(X) (3/\eps-2/3)&
X\lambda & \alpha_i C_i(G)(4/\eps-1)\\
\widetilde{\tilde{X}\tilde{X}}&\alpha_j C_j(X) \frac{8}{3}(-2/\eps-1)&
\tilde{X}\lambda & \alpha_i C_i(G)(1/\eps-1)\\
\widetilde{\tilde{X}\tilde{X}}{}'&
\alpha_j C_j(X) (\frac{5}{3}\frac{1}{\eps}+\frac{4}{3}) &
\lambda\lambda& \alpha_i C_i(G)(-4/\epsIR+4) \\  \hline
\end{array}$$
\caption{\em Contributions $\delta M_i^\Gamma=M_i^{(1)} V^\Gamma/4\pi$
of the single graphs $\Gamma$ in the limit $F\ll M^2$.}
\end{table}

\begin{figure}[t]\setlength{\unitlength}{1in}\begin{center}
\begin{picture}(5,5)
\ifMac
{\put(0,-0.3){\special{picture gra}}}
{\put(0,-0.3){\includegraphics{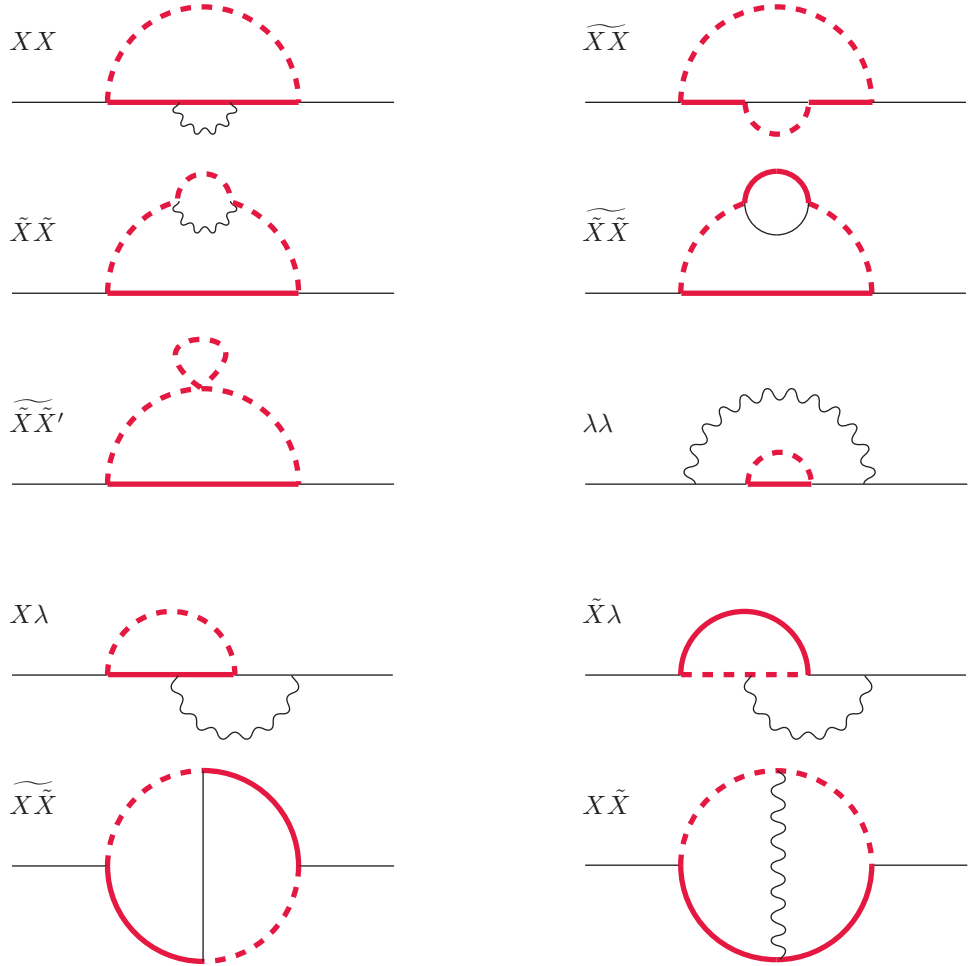}}}
\put(0,4.5){$XX$}             \put(3,4.5){$\widetilde{XX}$}
\put(0,3.5){$\tilde{X}\tilde{X}$} \put(3,3.5){$\widetilde{\tilde{X}\tilde{X}}$}

\put(0,2.5){$\widetilde{\tilde{X}\tilde{X}}{}'$}\put(3,2.5){$\lambda\lambda$}
\put(0,1.5){$X\lambda$}           \put(3,1.5){$\tilde{X}\lambda$}

\put(0,0.5){$\widetilde{X\tilde{X}}$}     \put(3,0.5){$X\tilde{X}$}
\end{picture}
\vspace{0.5cm}
\caption[SP]{\em The diagrams that contribute to the $NLO$ gauge corrections
to gauge mediated gaugino masses.
The thick continuous (dashed) lines represent the fermionic (bosonic) messengers.
The thin wavy (continuous) lines represent the gauge bosons (gauginos).
\label{fig:gra}}
\end{center}\end{figure}

\subsection{Computation in the full theory}\label{full}
The NLO correction to the pole gaugino masses are given by the
ten two-loop diagrams shown in figure~1 and some one-loop renormalization factor.
It is convenient to separate the renormalization factors
due to the light MSSM loops
from the ones due to messenger loops.
We write the various contributions to the pole gaugino masses as
\begin{equation}
M_i|^{\rm full}_{\rm pole}= \hM_i^{(1)}+\sum_\Gamma \delta M_i^\Gamma,\qquad
\hM_i^{(1)}\equiv\sum_n \hM_{in}^{(1)}=\frac{\alpha_i}{4\pi}
\sum_n \frac{F_n}{M_n}T_i(R_n) \hat{g}_1(x_n)=M_i^{(1)}+\Ord(\eps)
\end{equation}
where $x_n=F_n/M_n^2$ and the sum $\sum_n$ extends over all the messengers.
The one-loop function $\hat{g}_1(x)\equiv g_1(x)+\eps~\delta g_1(x)+\Ord(\eps^2)$ is~\cite{g1}
\begin{equation}\label{eq:g1}
\hat{g}_1(x) = \frac{1+x}{x^2}\ln(1+x)
\left\{1+\eps\,\big[1-\frac{\ln(1+x)}{2}\big]\right\}\big(1+\eps \ln\frac{\mub^2}{M^2}\big) +
(x\to -x)+\Ord(\eps^2)=
1+\frac{x^2}{6}+\Ord(x^4,\eps)
\end{equation}
All parameters are unrenormalized (`bare').
Here we list all the contributions to the pole gaugino masses.
\begin{itemize}
\item The contribution given by the sum of the {\bf two loop diagrams}
of fig.~\ref{fig:gra}. They give
\begin{eqnarray}\nonumber
\delta M_i^{\rm 2~loop} &=&\frac{\alpha_i}{4\pi} \sum_n
\frac{F_n}{M_n} T_i(R_n)\Bigg\{
\frac{\alpha_i}{4\pi}C_i(G)
\left[\hat{g}_1(x_n)(\frac{4}{\epsUV}-\frac{4}{\epsIR}+g_{\rm IR})+2g_2(x_n)\right]+\\
&&+\sum_{j=1}^3\frac{\alpha_{j}}{4\pi}C_{j}(X_n)
\left[g_C(x_n)-4g_1(x_n)-\frac{2x_n \hat{g}'_1(x_n)}{\epsUV}\right]\Bigg\}
\left(\frac{\mub^2}{M_n^2}\right)^{\!\!\eps}\label{eq:dM2loop}
\end{eqnarray}
where $\hat{g}'_1(x)$ is the derivative of $\hat{g}_1(x)$
with respect to $x$, and the functions $g_2$ and $g_C$ are
\begin{eqnsystem}{sys:g2gC}
g_2(x)&=&\frac{\ln(1+x)}{2x^2}\big[2(1+x)+\ln(1-x)+(2x+2+1/x)\ln(1+x)\big]+\\
\nonumber
&&+\frac{1}{x}\big[\Li(\frac{2x}{1+x})-2\Li(x)\big]+(x\to-x)\\
g_C(x)&=&
\frac{\ln(1+x)}{x^2}\big[8+6x+\ln(1-x) +(1-x+2/x)\ln(1+x)\big]+(x\to-x)
\end{eqnsystem}
In the limit $F\ll M^2$ ($x\to0$) $g_1(0)=g_2(0)=1$ and $g_C(0)=0$.
We have denoted as $1/\epsUV$ an $1/\eps$ ultraviolet (UV) pole,
and as $1/\epsIR$ a pole of infrared (IR) origin.
\lascia{The bilogarithmic function, $\Li$, cancels out from $g_C$.
The first five graphs (the ones with a 1~loop insertion in the messenger propagators)
contribute only to $g_C$.
Their sum contains an ultraviolet pole proportional to $(x g_1(x))'$.
The divergences present in the remaining graphs are all proportional to $g_1(x)$.
Including the two $X\tilde{X}$-type graphs
(the ones that contribute both to $g_2$ and to $g_C$),
the divergence in the last term of eq.\eq{dM2loop} becomes proportional to $x g'_1(x)$.
This divergence corresponds to a renormalization of $x=F/M^2$ and vanishes as $x\to 0$.}
In all the graphs it is possible to set the external gaugino momentum to zero,
except in the infrared divergent $\lambda\lambda$ graph of fig.~\ref{fig:gra}.
It is convenient to split it into a part computed with zero external momentum,
that contributes to $g_2$, plus the remainder,
that gives the $g_{\rm IR}$ term in eq.\eq{dM2loop}.
The value of $g_{\rm IR}$ coincides with the `non na\"{\i}ve part'
of the `asymptotic expansion'\footnote{
The general technique of asymptotic expansions of Feynman diagrams
is described in~\cite{HME};
a much simpler discussion, sufficient for the purposes of this computation,
can be found in~\cite{bsg}, where an accurate distinction between
UV and IR divergences is made.} in the external gaugino momentum,
and can be seen as the contribution of the $\lambda\lambda$ graph
with the heavy messenger loop contracted to a point.
This technical detail is useful, because a corresponding one-loop diagram
gives the same contribution to the effective theory, so that
we do not need to compute  $g_{\rm IR}$.

\item {\bf On-shell renormalization of the gaugino wave function}:
the renormalized gaugino field is
$\lambda_i=(1+z_i+{z}'_i)^{1/2}\cdot\lambda_i|_{\rm bare}$ with
$$z'_i=\frac{\alpha_i}{4\pi}\sum_n T_i(R_n)\left\{\frac{1}{\eps}+\ln\frac{\mub^2}{M_n^2}+
\frac{x_n^2 + (1-x_n^2) [\ln(1 - x_n) +\ln(1+x_n)]}{2x_n^2}\right\}.$$
We have separated this correction into two parts:
$z$ due to all the MSSM particles,
and ${z}'$ due to messenger loops only.
The corresponding corrections to the gaugino masses,
$\delta M_i^{z}+\delta M_i^{{z}'}$,
are obtained expressing the LO result in terms of the renormalized field $\lambda_i$:
$\delta M_i^{{z}'}=-\hat{M}_i^{(1)}\cdot {z}'_i$.
We do not need to specify the MSSM part because it is the same in both versions of
the theory.

\item {\bf Renormalization of the gauge couplings}:
we choose to express the bare gauge couplings of the {\em full} theory,
$g_i|_{\rm bare}$, as function of the quantum-corrected
gauge couplings in the {\em effective}
theory renormalized in the $\DRbar$ scheme, $g_i(\mub)$.
Defining $\alpha_i\equiv g_i^2/4\pi$ and including
the messenger thresholds effects we find
$$\alpha_i(\mub)=\alpha_i|_{\rm bare}-\delta\alpha_i-{\delta}'\alpha_i,\qquad
\frac{{\delta}'\alpha_i}{\alpha_i}=\frac{\alpha_i}{4\pi} \sum_n T_i(R_n)
\left\{\frac{1}{\eps}+\ln\frac{\mub^2}{M_n^2}
-\frac{\ln(1 - x_n) +\ln(1+x_n)}{6}\right\}.$$
We have separated the correction to $\alpha_i$ into two parts:
$\delta\alpha$ due to all the MSSM particles,
and ${\delta}'\alpha$ due to messenger loops only.
The corresponding corrections to the gaugino masses,
$\delta M_i^{\delta\alpha}+\delta M_i^{{\delta}'\alpha}$,
are obtained expressing the LO result in terms of the renormalized MSSM
gauge coupling $\alpha_i(\mub)$:
$\delta M_i^{{\delta}'\alpha}=\hat{M}_i^{(1)}\cdot {\delta}'\alpha_i/\alpha_i$.
We do not need to specify the MSSM part because it is the same in both versions of
the theory.
Notice that the messenger contribution
$\delta M_i^{{\delta}'\alpha}+\delta M_i^{{z}'}$
is UV finite.

\item {\bf Renormalization of $F$ and $M$.}
We employ $\DRbar$ renormalization: $F$ and $M$ are defined as
their bare values plus the pole parts of their quantum corrections.
As a consequence of the non-renormalization theorem,
the parameters $M_n$ and $F_n$ renormalize in the same way:
$$\{F_n,M_n\}=\big(1-\sum_j \frac{\alpha_j}{4\pi}2C_j(X_n)\big)\{F_n,M_n\}_{\rm bare} $$
The corrections to the gaugino masses
are obtained expressing the bare parameters in terms of the
renormalized ones in the one loop result:
\begin{equation}\label{eq:dMiF,M}
\delta M_i^{F,M}=\frac{\alpha_i}{4\pi}\sum_n  \frac{F_n}{M_n} T_i(R_n)
 \sum_j \frac{\alpha_j}{4\pi} C_j(X_n)
\left\{4g_1(x_n)+\frac{2x_n \hat{g}'_1(x_n)}{\epsUV}\right\} .
\end{equation}
The first term of\eq{dMiF,M} derives from $\ln\mub^2/M^2$ in\eq{g1} and, for $x=0$,
cancels the NLO correction to $M_i$ proportional to $\alpha_j$
produced by the two-loop diagrams.
\end{itemize}
As an aside remark, it could be of interest to know that the NLO
squared pole mass of
the lightest scalar messenger, $M_{n-}^2$, is 
\begin{eqnarray*}
\frac{M_{n-}^2}{M_n^2-F_n} &=&1-\sum_j \frac{\alpha_j}{4\pi}C_j(X_n)
\bigg\{2(x_n-2)(2+\ln\frac{\mub^2}{M_n^2})+\\
&&+
\frac{2x_n -3 - 3{x_n^2}}{x_n-1} \ln(1-x_n)+ 
\frac{2x_n^2}{x_n-1}\ln x_n-(1+x_n)\ln(1+x_n)\bigg\}
\end{eqnarray*}
when expressed in terms of $\DRbar$ parameters.

\begin{table}[t]{\small
$$\begin{array}{|c|ccc|ccc|}\hline
\hbox{rep.}    &C_1   &C_2  &C_3  & T_1 & T_2 & T_3\\ \hline\hline
d\oplus\bar{d} & 2/15 & 0   & 8/3 & 2/5 & 0   & 1 \\
L\oplus\bar{L} & 3/10 & 3/2 & 0   & 3/5 & 1   & 0 \\ \hline
Q\oplus\bar{Q} & 1/30 & 3/2 & 8/3 & 1/5 & 3   & 2 \\
u\oplus\bar{u} & 8/15 & 0   & 8/3 & 8/5 & 0   & 1 \\
e\oplus\bar{e} & 6/5  & 0   & 0   & 6/5 & 0   & 0 \\ \hline
\end{array}$$}
\caption{\em Values of the group factors for the $G_{\rm SM}$
fragments of the $5\oplus\bar{5}$ and
$10\oplus\overline{10}$ $\SU(5)$ representations.\label{tab:CT}}
\end{table}

\subsection{Computation in the effective theory}\label{eff}
In the effective theory we have to compute the pole gaugino masses in terms
of the coefficients of the running gaugino mass term operator,
$-M_i (\lambda_i\lambda_i+\hc)$,
expanded as a series in the gauge couplings:
\begin{equation}\label{eq:M1M2}
M_i=\frac{\alpha_i}{4\pi} M^{(1)}_i +
\frac{\alpha_i}{4\pi}\frac{\alpha_j}{4\pi} M^{(2)}_{ij}+\cdots
\end{equation}
where $M^{(1)}_i$ are the known LO coefficients, and
$M^{(2)}_{ij}$ are the NLO coefficients that we want ultimately to extract.
The pole gaugino masses at $\Ord(\alpha^2)$ order are given by
\begin{itemize}

\item the contribution from the renormalization of the gaugino wave function.
This coincides with the $\delta M_i^{z}$ correction present also in the full theory.

\item the contribution from the renormalization of the gauge couplings.
This coincides with the $\delta M_i^{\delta\alpha}$ correction present also in the full theory.

\item  a one loop diagram
(gauge correction to the gaugino propagator).
As said, it is not difficult to see that it gives the same contribution
of the asymptotic expansion in the external gaugino momentum
of the two-loop $\lambda\lambda$ diagram of fig.~\ref{fig:gra}.
\end{itemize}
More in detail the pole gaugino masses in the effective theory are
\begin{equation}\label{eq:Meff}
M_i|_{\rm pole}^{\rm eff}=\frac{\alpha_i}{4\pi} M^{(1)}_i
\left\{1+\frac{\alpha_i}{4\pi}C_i(G)
(\frac{4}{\epsUV}-\frac{4}{\epsIR}+g_{\rm IR})\right\}+
\frac{\alpha_i}{4\pi}\frac{\alpha_j}{4\pi}
M^{(2)}_{ij}+\delta M_i^z + \delta  M_i^{\delta\alpha}.
\end{equation}
A further simplification occurs.
The sum of the three effective-theory
quantum corrections, all proportional to $M^{(1)}_i$, is both infrared
and ultraviolet convergent
(because the combination $M_i/\alpha_i $ is RGE-invariant at one loop).
For this reason we do not need to worry about the $\Ord(\eps)$ terms that
distinguish $\hM^{(1)}_i$ from $M^{(1)}_i$.

\subsection{Matching}\label{match}
The matching procedure is particularly simple:
the running gaugino masses in the effective theory at NLO order are simply given
by the full theory result, omitting those quantum corrections that are
present also in the effective theory result, eq.\eq{Meff}.
The MSSM running gaugino masses at NLO order are
\begin{equation}\label{eq:MiNLO}
\begin{array}{rl}\Blue
M_i =&\displaystyle \frac{\alpha_i}{4\pi} \sum_n
\frac{F_n}{M_n} T_i(R_n)\Bigg\{g_1(x_n)+
\frac{\alpha_i}{4\pi}\left[2C_i(G) g_2(x_n)+\sum_m T_i(R_m) g_T(x_m)\right]+\\
&\displaystyle
+\sum_{j=1}^3\frac{\alpha_j}{4\pi}C_j(X_n) \left[g_C(x_n)-
2x_n g'_1(x_n)\ln\frac{\mub^2}{M_n^2}\right]
+\frac{\lambda_n^2}{(4\pi)^2} g_\lambda(x_n)\Bigg\}\Black.
\end{array}
\end{equation}
The parameters are renormalized as discussed in sec.~\ref{full}.
The last term is the effect
of a possible Yukawa coupling $\lambda_n \, S\,X_n\bar{X}_n$ in the superpotential,
where $S$ is a gauge singlet\footnote{The computation of Yukawa corrections,
that we do not present, involves one new feature:
a renormalization of $F_n/M_n$.
For simplicity, we have given the Yukawa correction
assuming that all the $F_n$ and $M_n$ are produced by
a vacuum expectation value of the superfield $S$.
In this case the new feature becomes an irrelevant common renormalization of $F_n/M_n$.
The corresponding effect in the final result, eq.\eq{MiNLO}, has been
absorbed in an appropriate (non $\DRbar$) renormalization of $F_n$.}.
The functions $g_1$, $g_2$ and $g_C$ have been defined in eq.s\eq{g1} and~(\ref{sys:g2gC}),
while $g_T$ and $g_\lambda$ are
\begin{eqnsystem}{sys:gTla}
g_T(x)&=&\frac{2x^2-3}{6x^2}\ln(1-x^2)-\frac{1}{2},\\
g_\lambda(x)&=&-\frac{\ln(1+x)}{2x^3}\left[2x(3+2x)-x\ln(1-x)+(5+3x)\ln(1+x)\right]
+\frac{x-4}{x^2}\Li(x)+(x\to-x).
\end{eqnsystem}
The functions $g_1$ and $g_2$ are normalized such that $g_1(0)=g_2(0)=1$,
while $g_C(0)=g_T(0)=g_\lambda(0)=0$.
So far we have assumed that $R_n=X_n\oplus\bar{X}_n$.
If there are also messenger fields $\Sigma$ in a real representation $R_\Sigma$,
the appropriate group factors are
$T(R_\Sigma)=T(\Sigma)$ and $C(X)\to C(\Sigma)$.

\medskip

In the limit $x_n\to 0$ the NLO prediction for the running gaugino
masses does not depend on the messenger spectrum
\begin{equation}
M_i(\mub) \stackrel{F\ll M^2}{=} \frac{\alpha_i(\mub)}{4\pi} \sum_n\frac{F_n}{M_n} T_i(R_n)
\left[1+\frac{\alpha_i}{4\pi}2C_i(G)\right]\qquad\hbox{at $\mub\sim M_n$}.
\end{equation}
We remember that $C_i(G)=\{0,2,3\}$.
This prediction for the three gaugino masses,
without knowing the values of the $F_n/M_n$ parameters,
is of interest only if their number is less than three.
This happens, for example, if the messengers lie in a $5\oplus\bar{5}$ representation of SU(5).

\begin{figure}[t]\setlength{\unitlength}{1cm}\begin{center}
\begin{picture}(16,8)
\ifMac{\put(0,0){\special{picture r10}}}{\put(0,0){\includegraphics{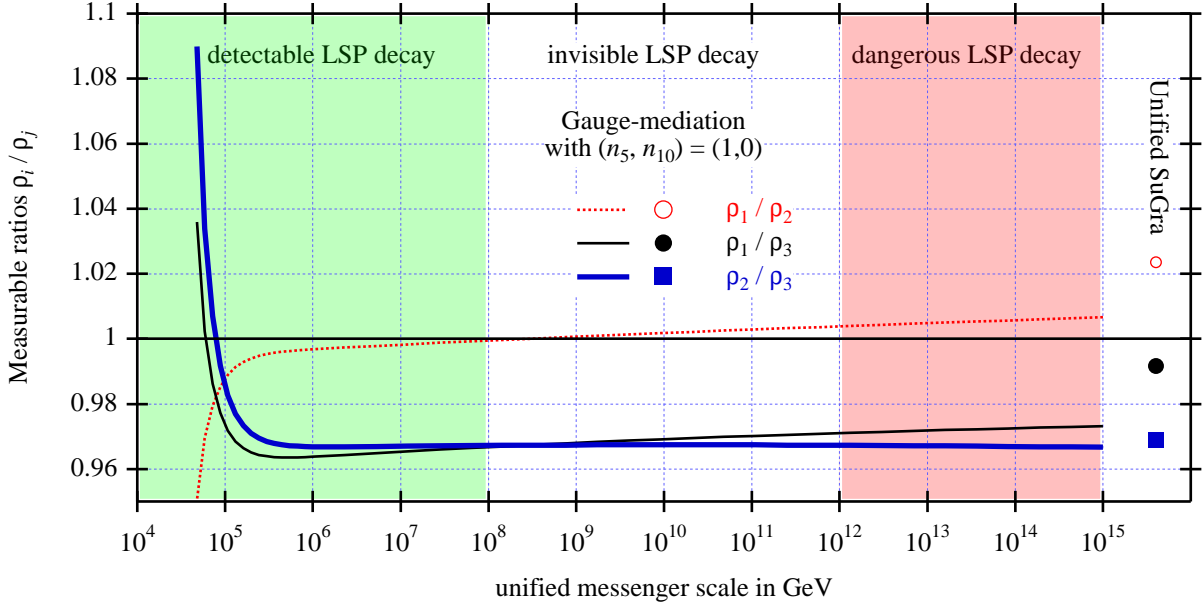}}}
\end{picture}
\caption[da3]{\em NLO predictions for the measurable ratios
$\rho_i(M_Z)/\rho_j(M_Z)$
($\rho_i\equiv M_i/\alpha_i$) in the minimal gauge mediated
model ($n_5=1$, $n_{10}=0$) for generic values of the unified messenger mass.
For comparison, we also show the prediction of unified supergravity models
(neglecting possible small GUT-scale effects).
\label{fig:rho10}}
\end{center}\end{figure}

\begin{figure}[t]\setlength{\unitlength}{1cm}\begin{center}
\begin{picture}(16,8)
\ifMac{\put(0,0){\special{picture rnm}}}{\put(0,0){\includegraphics{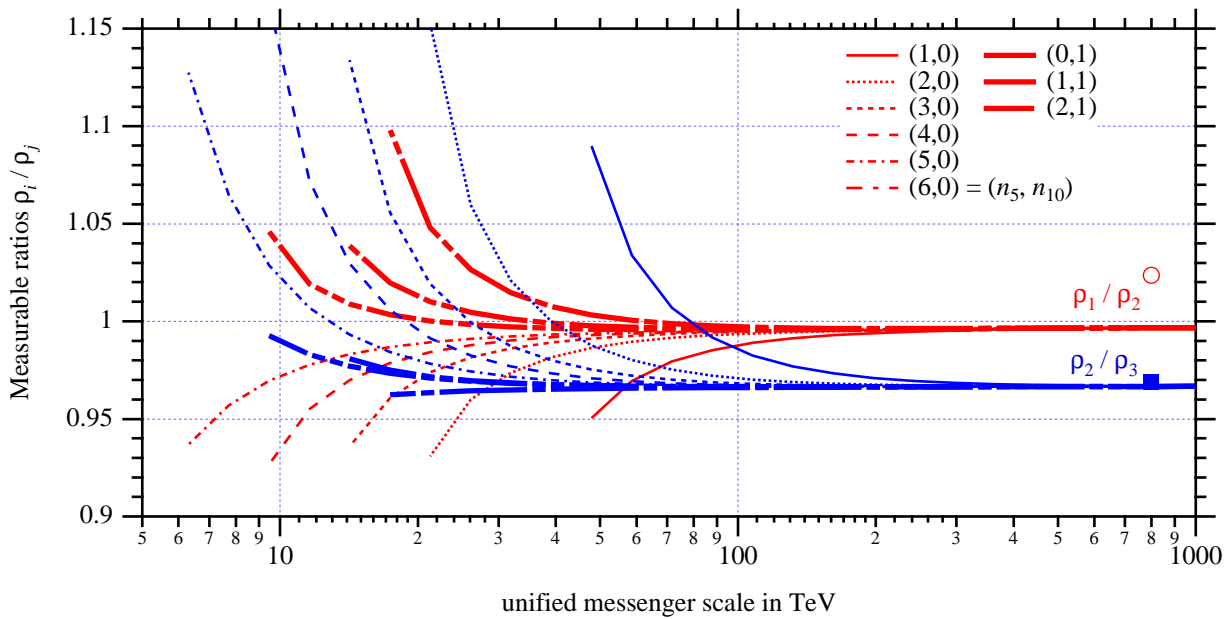}}}
\end{picture}
\caption[da3]{\em As in fig.~\ref{fig:rho10},
but with a `less minimal' messenger content.
\label{fig:rhoEtc}}
\end{center}\end{figure}

\begin{figure}[t]\setlength{\unitlength}{1cm}\begin{center}
\begin{picture}(16,8)
\ifMac{\put(0,0){\special{picture da3}}}{\put(0,0){\includegraphics{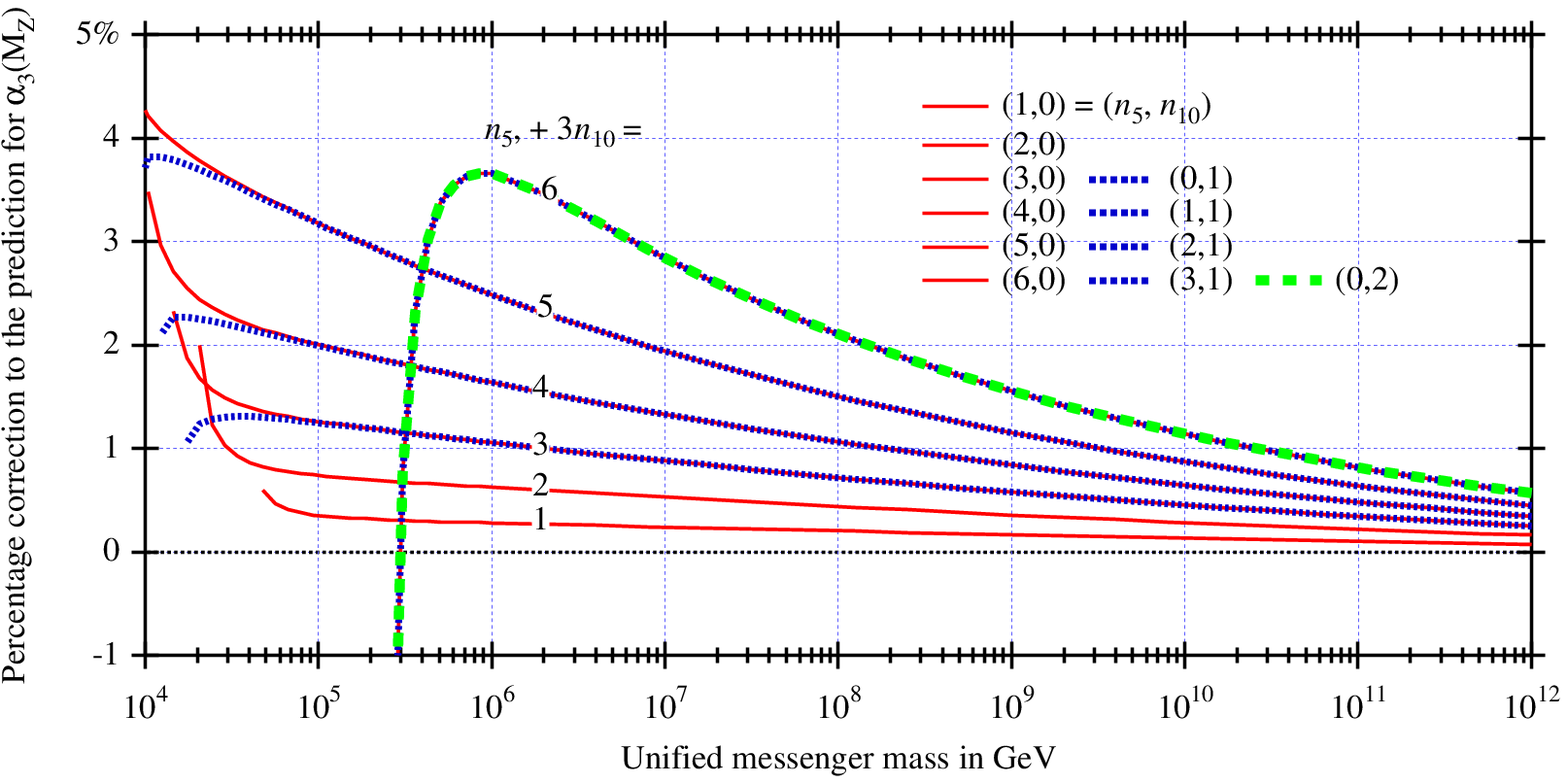}}}
\end{picture}
\caption[da3]{\em Percentage correction to the unification prediction for
$\alpha_3(M_Z)$ due to the presence of messengers
unified in the simplest representations of $\SU(5)$.
\label{fig:da3}}
\end{center}\end{figure}

\section{Predictions of unified messenger models}\label{fenomenologia}
We will show the NLO predictions in
models where the messenger spectrum satisfies unification relations.
These models are not only more predictive but also more appealing:
the successful unification of the gauge couplings is not destroyed and
a unified messenger spectrum helps in avoiding undesired
one-loop contributions to sfermion masses.
To be more specific we assume that the messengers fill $n_5$ copies
of $5\oplus\bar{5}$ and $n_{10}$ copies of $10\oplus\overline{10}$ representations
of the unified group SU(5),
so that the messenger contribution
to the one loop coefficient of the gauge $\beta$ functions is $T_i(R)=n_5+3n_{10}$.
We assume that the messenger mass parameters $M_n$ and $F_n$ arise
from the vacuum expectation value of one SU(5)-singlet field $S$ coupled
to the messengers via Yukawa interactions $\lambda_n\,S\,X_n\bar{X}_n$.
Imposing the unification relations
the running $\DRbar$ mass parameters at $\mub\sim M_n$ are thus obtained with NLO precision
via two-loop RGE evolution from $M_{\rm GUT}$ down to $\mub$:
$$x_n(\mub)=\frac{F_n}{M_n^2}={\cal U}_n(M_{\rm GUT}\to \mub) [x_N(M_{\rm GUT})+\delta_n x_N]
\qquad\hbox{for messengers $R_n$ unified in $R_N$}$$
where $\delta_n x_N$ represent unknown one-loop threshold effects at the unified scale,
that we will neglect.
`Reasonable' threshold effects give small corrections also when $x\sim1$.
We also neglect NLO Yukawa corrections, possibly relevant when $x\sim 1$.
The NLO RGE equations for $M_n$ and $F_n$ are given in appendix~B
(the combination $F_n/M_n$ is RGE-invariant, and is not corrected by threshold effects).
Already at LO,
the RGE evolution of the messenger spectrum gives
corrections of relative order $\Ord(x^2 \alpha\ln M_{\rm GUT}/M)$
to the relations $M_i\propto \alpha_i$:
for light messengers
the leptonic messengers are approximately $2$ times lighter than the hadronic ones.
This explains the larger effect present
for light messengers ($x_N\sim 1$)\footnote{
Values of $x_N>1$ (negative squared tree-level masses for scalar messengers)
give an acceptable spectrum of physical messenger masses ($x_n(\mub)<1$),
leaving only charge and/or colour breaking (CCB) minima
at very large field values.
This situation is not forbidden.
On the contrary, non-dangerous CCB minima
are quite generic in gauge mediated models~\cite{GMFT}.}.

Fig.~\ref{fig:rho10} shows the prediction of the minimal model with $(n_5,n_{10})=(1,0)$
for the measurable ratios $\rho_i/\rho_j$
(the experimental errors on the gauge couplings
negligibly affect the predictions for the $\rho_i(M_Z)\equiv M_i(M_Z)/\alpha_i(M_Z)$).
In this figure we have considered the whole range of possible messenger masses,
distinguishing the smaller values of the messenger mass for which the lightest supersymmetric
particle (LSP) decays before escaping detection,
from the higher values for which the LSP decay is so slow
(in absence of $R$-parity breaking)
that destroys the nucleosynthesys products~\cite{DDGR}.

In fig.~\ref{fig:rhoEtc} we consider models with more than a single unified messenger.
For the sake of illustration we have combined their contributions assuming that
all the messengers have the same unified $M_n$ and $F_n$.
For values of $M$ higher than the ones considered in fig.~\ref{fig:rhoEtc}
($x\ll1$)
all neglected NLO terms are completely irrelevant,
the prediction does not depend on the messenger content,
and is the same as in fig.~\ref{fig:rho10}.
In all plots we have fixed the $F$-terms requiring that
the running gluino mass be $M_3(M_Z)=500\GeV$.
Any other reasonable value of the gluino mass gives the same prediction for $\rho_i$.
We remember that we
have {\em not} included the one-loop corrections at the electroweak scale to $M_i$,
that depend on unmeasured (but measurable)
and unpredicted parameters (mainly the $\mu$-term).
The error on these predictions, due to remaining
NNLO effects, is estimated to be at the {\em per-mille} level,
much smaller than the expected experimental error
on the gaugino masses.

\smallskip

In fig.s~\ref{fig:rho10}, \ref{fig:rhoEtc} we
have also plotted the corresponding NLO prediction of unified supergravity models,
without including unknown possible GUT-scale corrections.
The unification relation $\rho_i\propto \One$ is infact only corrected at the $\%$ level by
GUT-scale threshold~\cite{SoglieMi} and gravitational~\cite{GravSmear} effects,
that could instead give much larger corrections to
the unification relations for the $\alpha_i$ and for the $M_i$.
The RGE contribution from the top $A$ term,
driven towards its IR fixed point value, $A_t(M_Z)\approx2M_2$,
is also numerically negligible~\cite{RGEM}.
The same can be said for the bottom and $\tau$ contributions,
that remain negligible also if $\tan\beta$ is large.

\medskip

At this point it is also interesting to discuss the NLO correction that the
presence of messenger fields gives to the {\em unification prediction for $\alpha_3(M_Z)$.}
The percentage correction to $\alpha_3(M_Z)$ is plotted in fig.~\ref{fig:da3}
for models with different messenger content, assuming that
all different messengers have a common unified value of the $M$ and $F$ parameters
(if $F\ll M^2$ it is only necessary to assume that the various $M_n$
have the same order of magnitude).
If $n_5+3n_{10}>5$ too light messengers give a non-perturbative value
of the unified gauge coupling.

It is interesting to see more in detail why this correction
is much smaller than its na\"{\i}ve expectation
and why it exhibits some curious property.
For given values of the unification scale and of the
unified gauge coupling, the low energy gauge couplings 
receive two contributions due to the presence of messengers:
$\delta\alpha_i^{-1}(M_Z)=\delta_i^{\rm th}+\delta_i^{\rm RGE}$.
The first contribution, $\delta_i^{\rm th}$, is due to messenger thresholds
(gauge corrections distort the unified messenger spectrum);
the second one is due to messenger corrections to the running of the gauge couplings
(we can reabsorb the one-loop contribution in the definition of the
unified gauge coupling, and consider only the two-loop contribution).
In a sufficiently accurate approximation the two corrections are
\begin{eqnsystem}{sys:delta}
\delta_i^{\rm RGE}&=&\frac{1}{4\pi}\sum_{j} \left\{ b^{(2)}_{ij}\ln r_j-
b^{(2)\rm MSSM}_{ij}\ln r_j^{\rm MSSM}\right\}\\
\delta_i^{\rm th} &=& \frac{1}{4\pi} \sum_n\sum_{p\in R_n} b^p_i\ln\frac{\mub^2}{M_p^2}
\stackrel{F\ll M^2}{=}-
\frac{1}{4\pi} \sum_j\sum_n 4T_i(R_n) C_j(X_n)\ln r_j\label{eq:dith}
\end{eqnsystem}
where $\sum_p$ extends over all the fermionic and scalar messengers,
$b^p_i$ is the contribution of a given messenger to
$b_i^{(1)}=b_i^{\rm MSSM}+\sum_p b_i^p$,
and $b_i^{(1)}$ and $b_{ij}^{(2)}$ are the one and two-loop coefficients
of the gauge $\beta$-functions in presence of messengers
(explicitly given in appendix~B).
We define $r_i\equiv (1-b_i \alpha_i(\mub)/(4\pi) \ln M_{\rm GUT}^2/\mub^2)^{-1/b_i}>1$
while $r_i^{\rm MSSM}$ is its value without messengers.
The overall correction to the unification prediction for the
strong coupling constant is
\begin{equation}
\delta^{\rm mess}\alpha_3(M_Z)=\delta^{\rm th}\alpha_3+\delta^{\rm RGE}\alpha_3=
-\alpha_3^2(M_Z) (\delta_i^{\rm th}+\delta_i^{\rm RGE})\Pi_i 
\end{equation}
where $\Pi_i=\{5/7,-12/7,1\}$.
The two corrections, $\delta^{\rm th}\alpha_3$ and $\delta^{\rm RGE}\alpha_3$
can be quite large ($\pm \Ord(0\div 20)\%$) and depend separately on $n_5$ and on $n_{10}$.
However the sum of the two contributions, plotted in fig.~\ref{fig:da3},
is {\em much smaller}, typically positive,
$\delta\alpha_3=(0\div 3)\%$, and depends only on the correction to the
one-loop $\beta$-function coefficient $b^{\rm mess}_i=n_5+3 n_{10}$
(this is not true in the limit $x\sim1$, where supersymmetry-breaking effects become relevant).
This cancellation can be seen summing the expression for
$\delta_i^{\rm th}$ in the limit $F\ll M$, eq.\eq{dith}
(in which we have inserted the messenger masses $M_n$ obtained via one-loop
RGE evolution), with the RGE correction
(in which we insert the general values of the two-loop $\beta$-function coefficients,
written in eq.\eq{bij}):
\begin{equation}
\delta_i^{\rm mess}=\delta_i^{\rm th}+\delta_i^{\rm RGE}
\stackrel{F\ll M^2}{=} \frac{1}{4\pi}\left\{
2C(G_i)b^{\rm mess}_i \ln r_i+
b_{ij}^{(2)\rm MSSM}\ln\frac{r_j}{r_j^{\rm MSSM}}\right\}.
\end{equation}
Corrections due to possible
messenger Yukawa couplings would cancel out.
These cancellations reproduce the exact result found by
J.\ Hisano and M.\ Shifman in~\cite{hol,RGEMhol}
working in toy models with the holomorphic
supersymmetric gauge couplings.
As shown in~\cite{MiHol} this same reason is at the basis of the analogous cancellation
between RGE and threshold effects encountered in our NLO computation
of gaugino masses at $F\ll M$.

\section{Conclusion}
We have computed the next-to-leading order corrections to
gaugino masses in gauge-mediated models
for generic values of the  messenger masses $M$.
In unified messenger models there are up to $10\%$ corrections to the
unification-like relations $M_i(\mub)\propto \alpha_i(\mub)$
between the running gaugino masses and the gauge couplings, but only if the
messengers are strongly splitted by supersymmetry breaking.
If instead $M>100$~TeV there are only small (few \%) corrections
to the leading-order approximation $M_i\propto \alpha_i$,
as shown in fig.s~\ref{fig:rho10} and~\ref{fig:rhoEtc}.

We have also studied the messenger corrections to gauge coupling unification.
As a result of cancellations, dictated by supersymmetry,
between large RGE and threshold corrections
the predicted value of the strong coupling constant is typically only
negligibly increased, as shown in fig~\ref{fig:da3}.

\smallskip

In the limit $M^2\gg F$ (heavy messengers) the same NLO prediction for gaugino masses,
together with NLO predictions for sfermion masses,
can be obtained~\cite{MiHol} combining the techniques described in~\cite{GR}
and~\cite{hol,RGEMhol}.
If $F\sim M^2$ it is more difficult to obtain
a NLO prediction of sfermion masses; however the LO results~\cite{g1} show that
the effects of large supersymmetry breaking in the messenger spectrum
are much less relevant in the sfermion sector than in the gaugino sector.

\small

\paragraph{Acknowledgements}
We are grateful to Riccardo Barbieri
for many discussions, and to Riccardo Rattazzi for having
pointed out a missing term (!) in the renormalization of our result.

\appendix
\setcounter{equation}{0}
\renewcommand{\theequation}{\thesection.\arabic{equation}}

\section{Relevant Lagrangian in quadri-spinor notation}
We consider a theory with messenger chiral superfields
$\Phi_L$ and $\Phi_R$ in self-conjugate complex
representations of the SM gauge group
(with generators $T$ and $-T^T$).
In presence of the superpotential
$$W=\int d^2\theta~\lambda S\, \Phi_L \Phi_R,\qquad
\lambda\md{\hat{S}}= M+\theta\theta F.$$
The messenger superfields $\Phi=A+\sqrt{2}\theta\psi+\cdots$
contain the following mass eigenstates:
messenger fermions $\psi_L$, $\psi_R$ with a Dirac mass $M$,
and pseudoscalar (scalar) messengers
$A_\pm\equiv(A_L\pm A^*_R)/\sqrt{2}$ with mass
$M_\pm^2=M^2\pm F\equiv M^2(1\pm x)$.
The supersymmetric gaugino Lagrangian is
$$
{\cal L}_\lambda  =  \bar{\lambda} \bar{\sigma}^\mu iD_\mu\lambda -
\frac{m}{2}(\lambda\lambda + \bar{\lambda}\bar{\lambda})-
\sqrt{2}g(A^\dagger_L T \psi_L\lambda+\bar{\lambda} \bar{\psi}_L T A_L-
\lambda\psi_R^T T A_R^*-A_R^T T \bar{\psi}_R^T \bar{\lambda})
$$
where $D_\mu$
is the standard gauge-covariant derivative
and $\lambda$, $\psi_L$ and $\psi_R$ are Weyl fermions with the same chirality.

We want to rewrite the Lagrangian in terms of mass eigenstates.
In order to employ our {\tt Mathematica}~\cite{Math} code for analytic computation
of Feynman graphs, we need to write the messenger fermions as Dirac
quadri-spinors $\Psi$ and the
gauginos $\lambda$
as Majorana spinors $\Lambda$:
$$\Psi\equiv {\psi_L\choose \bar{\psi}_R^T},\qquad
\Lambda={\lambda\choose \bar{\lambda}}=\Lambda^c\equiv C\bar{\Lambda}^T$$
where $C$ is the charge-conjugation matrix.
The gaugino Lagrangian becomes
\begin{equation}
{\cal L}_\lambda  =  \frac{1}{2}\bar{\Lambda} (i\Dsl -m)\Lambda-
g(A^\dagger_- T \bar{\Lambda}\Psi+
\bar{\Psi}\Lambda T A_-  -
A_+^\dagger \bar{\Lambda}\gamma_5 \Psi  +
\bar{\Psi}\gamma_5 \Lambda T A_+)
\end{equation}
and the $D$ terms become
$$D^a=A_L^* T^a A_L - A_R T^a A_R^* + \cdots=
A_+^* T^a A_- + A_-^* T^a A_+ + \cdots $$
The gaugino $\Lambda\bar{\Lambda}$ propagator is a standard fermion propagator.
It is possible to show that graphs
with $\Lambda\Lambda$, $\bar{\Lambda}\bar{\Lambda}$ and
$\bar{\Lambda}\Lambda$ propagators have the same value
of the standard `$\Lambda\bar{\Lambda}$',
by appropriately rewriting the vertices in terms of
charge-conjugated fields\footnote{Only the $\widetilde{X\tilde{X}}$ diagram
of fig.~\ref{fig:gra},
that needs a Majorana gaugino propagator to be non-zero,
requires a more detailed treatment of the charge conjugation factors.}.
This shows that the gaugino can be treated as an ordinary fermion field,
but with symmetry factors computed like the ones of a {\em real} field.

In a general theory there will be several messenger pairs,
each one with its $M_n$, $F_n$ and $x_n\equiv F_n/M_n^2$.

\section{RGE evolution}
The necessary RGE can be read from the literature~\cite{RGEM}.
The RGE equations for gauge couplings and gaugino masses in the $\DRbar$ scheme are
\begin{eqnarray*}
\frac{d}{dt}\frac{1}{g_i^2}&=&b^{(1)}_i  +\frac{1}{(4\pi)^2} 
\left[\sum_jb^{(2)}_{ij} g_j^2 -\sum_a b^{(2)}_{ia} \lambda_a^2\right]\\
\frac{d}{dt}M_i&=&-g_i^2 b^{(1)}_i M_i +\frac{g_i^2}{(4\pi)^2}
\left[-\sum_j b^{(2)}_{ij} g_j^2 (M_i+M_j) + \sum_a b^{(2)}_{ia} \lambda_a^2 (M_i-A_a)\right]
\end{eqnarray*}
where $t(E)\equiv(4\pi)^{-2}\ln M_{\rm GUT}^2/E^2$ and $a$ runs over the
third generation particles, $a=\{t,b,\tau\}$.
In a general supersymmetric model with $\{\Phi\}$ matter superfields the coefficients are
\begin{equation}\label{eq:bij}
b_i^{(1)}=-3 C(G_i)+\sum_\Phi T_i(\Phi),\qquad
b_{ij}^{(2)}=2C(G_i) b_i^{(1)}\delta_{ij}+4\sum_\Phi T_i(\Phi)C_j(\Phi).
\end{equation}
In the models under consideration the field content is given by
the MSSM fields plus $n_5$ copies of $5\oplus\bar{5}$ and
$n_{10}$ copies of $10\oplus\overline{10}$
SU(5) messenger multiplets
(in the effective theory without messengers, the coefficients 
are obtained taking $n_5=n_{10}=0$).
The values of the coefficients are
$$
b_i^{(1)}=\pmatrix{33/5 \cr  1 \cr -3}+ (n_5+3n_{10})\pmatrix{1\cr1\cr1}\qquad
b_{ia}^{(2)}=\!\!\bordermatrix{a:&t&b&\tau\cr& 26/5 & 14/5 & 18/5\cr
&6 & 6 & 2\cr &4 & 4 &0}$$
$$
b_{ij}^{(2)}=\pmatrix{
{\frac{199}{25}} + {\frac{7\,{n_5}}{15}} + {\frac{23\,{n_{10}}}{5}}&
{\frac{27}{5}} + {\frac{9\,{n_5}}{5}} + {\frac{3\,{n_{10}}}{5}} & 
{\frac{88}{5}} + {\frac{32\,{n_5}}{15}} + {\frac{48\,{n_{10}}}{5}} \cr 
{\frac{9}{5}} + {\frac{3\,{n_5}}{5}} + {\frac{{n_{10}}}{5}} & 
25 + 7\,{n_5} + 21\,{n_{10}} & 24 + 16\,{n_{10}} \cr 
{\frac{11}{5}} + {\frac{4\,{n_5}}{15}} + {\frac{6\,{n_{10}}}{5}} &
9 + 6\,{n_{10}} & 14 + {\frac{34\,{n_5}}{3}} + 34\,{n_{10}} \cr  }.$$
In numerical computations it is useful to
employ the RGE for $\rho_i\equiv M_i/\alpha_i$, since it starts at two-loop order
\begin{equation}
\frac{d}{dt}\rho_i=-\frac{g_i^2}{4\pi}
\left[\sum_j b^{(2)}_{ij} g_j^2 M_j + \sum_a b^{(2)}_{ia} \lambda_a^2 A_a\right].
\end{equation}
When $F\sim M^2$ supersymmetry is {\em hardly\/} broken in the
effective theory below the messenger scale:
the couplings at the supergauge gaugino vertices differ
(by a numerically negligible amount) from the corresponding gauge couplings.
However in this case the messengers are light so that it
is sufficient to employ
the one loop RGE equations below the messenger scale.

Finally the gauge contribution to the
2~loop $\DRbar$ RGE equations
for the supersymmetric messenger masses $M_n$ and for the `$F$-terms' $F_n$ is
$$\frac{d}{dt}\ln M_n=\frac{d}{dt}\ln F_n=2C_i(X_n) g_i^2 -\frac{1}{(4\pi)^2}
\left\{g_i^2 g_{j}^2 \, 2C_i(X_n)\,2C_{j}(X_n)+
g_i^4\,2C_i(X_n) b_i^{(1)}\right\}.
$$
\lascia{The explicit values of the two-loop RGE coefficients are
\begin{eqnarray*}
b_{dij}^{(2)}&=&
\pmatrix{ {\frac{202}{225}} + {\frac{2{n_5}}{15}} + {\frac{2n_{10}}{5}}
   & 0 & {\frac{16}{45}} \cr 0 & 0 & 0 \cr {\frac{16}{45}} & 0 & 
  -{\frac{8}{9}} + {\frac{8{n_5}}{3}} + 8n_{10} \cr  }\\
b_{Lij}^{(2)}&=&\pmatrix{ {\frac{207}{100}} + {\frac{3n_5}{10}} + {\frac{9n_{10}}{10}}
   & {\frac{9}{20}} & 0 \cr {\frac{9}{20}} & 
  {\frac{15}{4}} + {\frac{3n_5}{2}} + {\frac{9n_{10}}{2}} & 0 \cr 0
   & 0 & 0 \cr  }\\
b_{Qij}^{(2)}&=&\pmatrix{ {\frac{199}{900}} + {\frac{n_5}{30}} + {\frac{n_{10}}{10}} & 
  {\frac{1}{20}} & {\frac{4}{45}} \cr {\frac{1}{20}} & 
  {\frac{15}{4}} + {\frac{3n_5}{2}} + {\frac{9n_{10}}{2}} & 4 \cr 
  {\frac{4}{45}} & 4 & -{\frac{8}{9}} + {\frac{8n_5}{3}} + 8n_{10}
   \cr  }\\
b_{uij}^{(2)}&=&\pmatrix{ {\frac{856}{225}} + {\frac{8n_5}{15}} + {\frac{8n_{10}}{5}}
   & 0 & {\frac{64}{45}} \cr 0 & 0 & 0 \cr {\frac{64}{45}} & 0 & 
  -{\frac{8}{9}} + {\frac{8n_5}{3}} + 8n_{10} \cr  }\\
b_{e11}^{(2)}&=&{\frac{234}{25}} + {\frac{6n_5}{5}} + {\frac{18n_{10}}{5}}\\
\end{eqnarray*}
All other elements of $b_{eij}^{(2)}$ vanish.}
\lascia{
We have not considered RGE corrections to $x_n$
due to possible Yukawa couplings in the messenger sector.
If $F\ll M^2$ such couplings are indeed suggested by realistic models~\cite{GMmodels,DDGR}:
but in this limit the gaugino masses do not depend on $x_n$.}

\small~

\end{document}